# Structural dynamics probed by high-coherence electron pulses

Armin Feist[1], Gero Storeck[1], Sascha Schäfer[2], and Claus Ropers[1]

[1] University of Göttingen, IV. Physical Institute, Göttingen, Germany

[2] Carl von Ossietzky Universität Oldenburg, Institute of Physics, Oldenburg, Germany

Ultrafast measurement technology provides essential contributions to our microscopic understanding of the properties and functions of solids and nanostructures. Atomic-scale vistas with ever-growing spatial and temporal resolution are offered by methods based on short pulses of x-rays and electrons. Time-resolved electron diffraction and microscopy are among the most powerful approaches to investigate non-equilibrium structural dynamics in excited matter. In this article, we discuss recent advances in ultrafast electron imaging enabled by significant improvements in the coherence of pulsed electron beams. Specifically, we review the development and first application of Ultrafast Low-Energy Electron Diffraction (ULEED) for the study of structural dynamics at surfaces, and discuss novel opportunities of Ultrafast Transmission Electron Microscopy (UTEM) facilitated by laser-triggered field emission sources. These and further developments will render coherent electron beams an essential component in the future of ultrafast nanoscale imaging.

**Keywords:** photoemission, phase transformation, nanoscale, laser-induced reaction, transmission electron microscopy (TEM).

**Introduction**

The physics of modern functional materials is governed by nanoscale heterogeneity and processes occurring on fast to ultrafast timescales. Observations with high spatial and temporal resolution promote a comprehensive understanding of microscopic couplings and correlations, ultimately facilitating an active control of properties, excitations and transformations. Ultrafast metrology, including time-resolved optical and photoelectron spectroscopy,[1] electron and x-ray diffraction[2–4] and spectroscopy, yields access to multiple degrees of freedom down to the femtosecond and attosecond range.

Structural dynamics in strongly heterogeneous systems, such as nanostructures, surfaces or monolayers, are of particular interest, but also pose great experimental challenges. In recent years, various techniques for ultrafast nanoscale probing were established, enabling, for example, time-resolved scanning near-field optical microscopy (SNOM)[5–7] and scanning tunneling microscopy (STM),[8–10] electron point-projection microscopy (PPM),[11–13] and lensless imaging using extreme ultraviolet radiation.[14–16]

Ultrashort electron pulses are ideally suited probes of materials structure and dynamics down to the atomic level.[17, 18] Ultrafast diffraction in transmission,[3, 17, 19–23] grazing incidence[24–26] and backscattering[27] geometries is now established. Nanoscale real-space imaging, on the other hand, is possible using ultrafast transmission[18, 28, 29] and scanning[30, 31] electron microscopy. These techniques are rapidly expanding their scope, driven by strong developments of high-brightness pulsed electron sources.[22, 32–37]

In this contribution, we discuss two complementary schemes to access ultrafast structural dynamics using electron pulses from localized photoelectron sources. Specifically, Ultrafast Low-Energy Electron Diffraction (ULEED) probes the orientation, symmetry and long-range order of materials with ultimate surface sensitivity, while Ultrafast Transmission Electron Microscopy (UTEM) translates atomic-scale imaging and local diffraction to the ultrafast domain. We will introduce these methods and will discuss selected examples of laser-induced structural dynamics probed by them.

**Electron pulses from nanoscale sources**

Following the principle of optical pump-probe techniques, ultrafast electron diffraction and microscopy use a stroboscopic approach, as illustrated in Figs. 1a and b: An ultrashort laser pulse (red) excites a TEM specimen or surface, respectively, and the structural response is probed by a delayed electron pulse (green) in transmission or backreflection. By repetitive sampling at varying delays, a time-dependent 'movie' of the induced dynamical process can be composed. The real-space and reciprocal space resolution of these techniques crucially depend on the quality of the employed electron beam.[38] As a figure-of-merit, the brightness $B \sim I/(\varepsilon_x \cdot \varepsilon_y)$ is proportional to the beam current I per occupied transverse phase space area, quantified by the beam emittances $\varepsilon_{x,y}$ (see Refs. [38] for details). Electron pulses are typically generated by laser-induced photoemission, and the intrinsic source emittance can thus be reduced by optimizing the photocathode work function and material,[39, 40] and by tailoring the wavelength[41] or focussing conditions.[35, 42] Following a different path, our laboratory and other groups are actively pursuing ultrafast laser-triggered electron emission from nanoscale tip-shaped photocathodes,[43–47] yielding high-coherence electron pulses[48–50] in tunable electrostatic emitter geometries.[21, 49, 51–53]

The low intrinsic source emittance and large degree of coherence provided by nanoscale photoemitters (cf. Figs. 1c and d) allows for highly collimated and tightly focused pulsed electron beams, which is instrumental to the two experimental approaches described below, ULEED and UTEM.

**ULEED: Ultrafast Low Energy Electron Diffraction**

We first present a novel approach to track ultrafast structural dynamics at surfaces and monolayer films. This work is motivated by the particular importance of low-dimensional systems, many of which exhibit unique electronic, structural and magnetic phases and excitations. Spontaneous symmetry breaking and various types of topological states are a hallmark of surface systems,[54] and both their optical control and ultrafast dynamics are an exciting field of study. Today, detailed snapshots of the evolution of electronic excitations at surfaces can be recorded by angle-resolved photoelectron emission spectroscopy.[1] In contrast, ultrafast access to the associated structural degrees of freedom has been much more restricted until recently.

Electron beams represent an ideal probe of surface structure by backscattering diffraction. In particular, diffraction with a surface sensitivity to the first few monolayers is obtained at very low normal momentum. In Reflection High-Energy Electron Diffraction (RHEED),[55] this is achieved by employing a grazing incidence geometry, and several ultrafast implementations of this technique have been successful.[24, 25] However, being a much more widely used method in surface science, Low-Energy Electron Diffraction (LEED) at normal incidence is arguably the preferred choice for obtaining quantitative information on the symmetry and long-range order of a surface.[56] In our laboratory, we have therefore undertaken the development of Ultrafast Low-Energy Electron Diffraction (ULEED), which has resulted in first demonstrations of observing ultrafast structural transitions in transmission[21] and backscattering diffraction.[27, 57]

The challenge in implementing ULEED is the need for ultrashort (picosecond or femtosecond) electron pulses at low energies. At the corresponding slow electron propagation speeds, an initially short electron pulse with some energy spread is substantially broadened until it reaches a sample. Minimizing the propagation distance from the photoelectron source to the sample, we realized an ultrafast LEED setup in a transmission geometry to study the dynamics of a polymer layer on free-standing graphene.[21] However, for the more generic case of backscattering, shorter source-sample distances require drastic downscaling of the electron gun to avoid shadowing. We addressed this issue by implementing two types of very compact photoelectron guns with diameters in the millimeter [Fig. 2b] and micrometer [Figs. 2c,d] range, respectively. Both geometries consist of an electrostatic lens assembly hosting a laser-driven nanotip electron source. Specifically, the micrometer scale electron gun [Fig. 2 c,d] is fabricated by optical lithography and focused ion beam etching.[57] An exemplary diffraction pattern and the temporal resolution of presently 1.4 ps[27] are shown in Figs. 2e-g. Notably, the high beam coherence facilitated by the localized electron source allows for very sharp diffraction spots (Fig. 2f). Here, higher-order satellite diffraction spots from a periodic lattice distortion in 1T-TaS$_2$[58] are clearly resolved. The combination of high momentum and high temporal resolution with ultimate surface sensitivity makes ULEED a versatile approach to study structural dynamics and ordering phenomena at surfaces, as illustrated in the following.

**Phase ordering of charge density waves resolved by ULEED**

We recently employed the ULEED capabilities to investigate optically induced structural phase transitions and the emergence of long-range order following an ultrafast quench.[27] Specifically, here, we address the transition between two charge-density wave (CDW) phases of 1T-TaS2, a layered transition metal dichalcogenide. This material is an intensely studied model system displaying various CDW phases[59] and metastable states,[60] the transitions between which were observed by time-resolved photoemission spectroscopy,[61–64] electron diffraction[65, 66] and x-ray diffraction.[67] At room temperature, the so-called "nearly commensurate" (NCP) phase of the CDW exhibits a periodic lattice distortion (PLD) that is locally commensurate to the atomic lattice, but globally forms a weakly disordered domain pattern (Fig. 3a). At elevated temperatures above 353 K, the structure transforms to an incommensurate phase (ICP), i.e., a "floating" phase not pinned to the lattice. In LEED, the associated PLDs lead to hexagonal arrangements of satellite diffraction spots (encircled in Fig. 3b) surrounding the lattice Bragg peaks (bright spots), with different rotation angles.

The transition between these two phases can be optically triggered, and ULEED allows us to closely trace the corresponding structural evolution in time. Figure 3b presents diffraction images recorded at different delay times $\Delta t$ of the electron pulse relative to the pump pulse. Specifically, at negative delays (Fig. 3b, left), we observe the unperturbed NCP, while at large positive delay times, the structure is found in the ICP (Fig. 3b, right), characterized by rotated satellite spots and the lack of higher order satellite peaks. A difference image highlighting the disappearance of the NCP (blue) and the appearance of the ICP (red) diffraction peaks is shown in the center of Fig. 3b. A time series of close-ups near one of the satellite peaks is displayed in Fig. 3c, while Fig. 3d plots the respective delay-dependent spot intensities of the NCP (blue symbols) and ICP (red symbols), integrated over a circular area-of-interest with a diameter of 0.12 Å-1 in k-space (shown in Fig. 3c). Whereas the suppression of the NCP order occurs within the first picosecond (compare also Ref. 66), the rise of the ICP peak proceeds over tens to few hundred picoseconds.

Facilitated by the high momentum resolution of the ULEED setup, a detailed spot profile analysis shows a diffraction peak narrowing during this process,[27] which implies an increase of spatial correlation length $L_{IC}$ of the ICP structure. Considering the instrument response function, a power law scaling in time is identified, i.e., $L_{IC} \sim t^m$ with $m \approx 0.5$ (Fig. 3e). Numerical simulations within a Ginzburg-Landau approach reproduce this scaling,[27] and indicate that the formation and annihilation of topological defects play an essential role in this phase-ordering mechanism. Specifically, the phase transition proceeds via quenching into a highly disordered nascent ICP state, which contains a significant density of amplitude and phase fluctuation modes of the CDW ("amplitudons" and "phasons"), but also rather persistent, topologically protected dislocation-type phase vortices in the order parameter, namely the charge density modulation. The presence of such dislocations, theoretically discussed by McMillan in 1975,[68] limits the CDW correlation length and thus broadens the diffraction peaks. The establishment of long-range order then proceeds via the mutual annihilation of these CDW dislocations with a density-dependent rate, resulting in the power-law scaling described above.

These direct observations of the phase-ordering kinetics in a charge density wave system illustrate the strength of spot-profile analysis using parallel-beam ultrafast electron diffraction. A wide variety of nanoscale ordering phenomena, in particular involving topological defects and domain structures, will be accessible using this approach. As for further developments of the ULEED technique, we believe that further miniaturization as well as active phase space manipulation of electron pulses by time-varying fields[32, 69, 70] and possibly aberration correction will all play a role in its enhancement and proliferation.

**UTEM: Ultrafast Transmission electron microscopy**

In the second part of this contribution, we present our work on the implementation of nanoscale photocathodes in ultrafast transmission electron microscopy (UTEM). The use of pulsed electron sources for time-resolved TEM was pioneered with different points of emphasis at the Technical University of Berlin,[28] Lawrence Livermore National Laboratory (LLNL),[29] and Caltech.[71] Ultrafast TEM (UTEM) is a stroboscopic variant that allows for simultaneous nm real-space and sub-ps temporal probing, and is currently pursued and developed further in a number of laboratories worldwide.[40, 50, 72–79] At present, several instrumental and conceptual challenges remain, with the development of brighter ultrafast electron sources as a central target. Until recently, flat photocathodes were exclusively utilized in time-resolved TEM. Employing high-brightness nanoscale tip-shaped photocathodes in UTEM, however, promises a number of advantages, and may transfer state-of-the-art coherent beam techniques, such as electron holography,[80, 81] advanced beam-shaping,[82] and nanoscale diffractive probing,[83] into the femtosecond regime.

At the University of Göttingen, we recently implemented an advanced UTEM instrument making use of sharp field-emitter tips as on-demand source of high-coherence ultrashort electron pulses[49, 73] (cf. Fig. 4 a,b). Specifically, we modified a commercial JEOL JEM 2100F field emission TEM, facilitating localized single photon photoemission from a laser-triggered Schottky field emitter[30, 84] (cf. Fig. 4 c) and synchronized optical sample excitation. Details on the instrument, a comprehensive beam characterization and prospects for future applications are outlined in Ref. *49*.

In the limit of isolated single-electron emission, we achieve record electron pulse properties, with down to 0.9-Å electron focal spot size, 0.6-eV spectral width and 200-fs pulse duration (cf. Fig. 4 d-f). Within the duty cycle of the pulsed electron beam, a peak brightness comparable to that of a conventional Schottky emitter is obtained, facilitating an ultralow transverse beam emittance of down to 1.7 pm·rad.[49] Besides nm-probing capability using a sharply focused electron spot, the corresponding high degree of transverse coherence enables time-resolved variants of high-resolution phase-contrast imaging, electron holography, differential phase contrast (DPC) STEM and diffraction from macro-molecular/mesoscopic structures.

Notably, the emitter can be operated in conventional Schottky mode, facilitating standard TEM applications and in-depth sample characterization with µA of emission current. In photoelectron-mode, the temporal structure of the beam is freely tunable from sub-ps pulses to continuous operation, allowing for optimizing the maximum beam current to the requirements of a specific experiment. In the following Section, we describe the application of this instrument to the spatially resolved probing of strain dynamics using ultrafast convergent beam electron diffraction (U-CBED).

**Local Diffractive Probing of Ultrafast Strain Dynamics in UTEM**

In recent years, high-frequency opto-phononic devices operating in the MHz to THz range, such as toroidal microcavities,[85] suspended membrane structures,[86] or tailored multilayers, have shown great potential for advanced metrology and sensing applications.[87–89] The characterization of these devices utilizing high-resolution and ultrafast optical spectroscopies has provided a deep understanding of their optical properties but only indirectly gave access to the their phononic degrees of freedom and are often limited to micrometer length scales. Ultrafast diffractive probing in UTEM offers unique capabilities to locally track the transient structural distortions in reciprocal space on femtosecond time scales,[90–92] with exciting opportunities such as the mapping of phase transitions in intense optical near-fields or the imaging of tailored nanoscale phononic wave fields.

As a first example of U-CBED with a high-coherence electron source, we investigated the light-induced phononic wave field at the edge of a single-crystalline graphite membrane (cf. Fig. 5a).[93] The structure is illuminated with femtosecond laser pulses (50-µm focal spot size, 800-nm central wavelength, 50-fs pulse duration, 16-mJ/cm² optical fluence), and we diffractively probe the local structural distortion employing 700-fs electron pulses focused to a 28-nm diameter. An exemplary CBED pattern is shown in Fig. 5b. The electron focus contains a broad distribution of incidence angles on the sample, which is visible in the pattern as a central intense disc. For certain incidence angles, Bragg scattering conditions are fulfilled, appearing as line-like features in the CBED pattern.[83] The angular position and scattering intensity encodes the local unit cell structure and lattice temperature.

In Fig. 5c, we display the temporal change of the (422) Bragg line for two probing positions, far from the membrane edge (upper panel) and at an edge distance of 500-nm (lower panel). In both cases, pronounced shifts and modulations of the Bragg line profiles are observed, but with a different temporal behavior for both probing positions.

A particular strength of ultrafast diffractive probing is the ability to obtain quantitative information on structural distortions. By considering the temporal change of the mean position of a larger number Bragg lines, we extract the dominant components of the local distortion tensor and thereby identify the character of the light-triggered phononic modes, as plotted in Fig. 5d. For both probing positions, optical excitation results in a breathing of the film thickness, with a characteristic frequency of about 18 GHz (Fig. 5d, inset) given by the round-trip time of a longitudinal acoustic wave travelling in the out-of-plane direction. Close to the membrane edge, additionally a shear-rotational mode is observed (Fig. 5d, lower panel, red curve) with a frequency corresponding to the sound velocity of a transverse acoustic mode. We located the origin of the shear mode excitation by mapping the spatio-temporal evolution of the distortion tensor. Whereas the film breathing mode (Fig. 5e (upper panel)), is found at all sample positions with an approximately constant amplitude and phase, the shear-rotational component (lower panel) exhibits a distinct spatial dependency: starting at the edge of the membrane, the first maximum of the shear-rotational mode amplitude shows a linear relation to the distance from the membrane edge. This peculiar coupling of nanoscale phononic modes was qualitatively reproduced in a continuum mechanics model which included the ultrafast heating of the graphite lattice.[93]

In addition to the average local structural distortion, the Bragg line profiles in ultrafast CBED contain detailed information on the strain distribution within the depth of the membrane. In particular, the transient splitting of line profiles visible in Fig. 5c (lower panel) are caused by the inhomogeneous structural deformation of the travelling strain waves bouncing between the two membrane faces. Such a rich experimental data may further elucidate the role of the ballistic transport of optically generated carriers and high frequency (thermal) phonons and their mode-specific coupling to an average thermal stress.

**Further Perspectives**

The development of ULEED and the use of laser-triggered field emitters in UTEM are both still in their early stages. However, the first experiments have already shown the great benefit of localized photoelectron sources in these applications. We expect that ULEED will serve as a unique and strong complement to photoemission spectroscopy in ultrafast surface science, elucidating the dynamics of tailored surface excitations and contributing to the understanding and control of low-dimensional structural phase transitions. UTEM, on the other hand, is an incredibly versatile experimental platform, which will continue to grow in terms of its range of applications from time-dependent versions of local electron diffraction, imaging and spectroscopy to phenomena and approaches not encountered in conventional electron microscopy, including quantum optics with free-electron beams or attosecond phenomena.[94, 95] We believe that coherent ultrafast electron sources will play an essential role in each of these aspects within the emerging field of ultrafast nanoscale imaging.

**Acknowledgments**

We gratefully acknowledge the collaboration with our co-authors on the works discussed here. This work was funded by the Deutsche Forschungsgemeinschaft (DFG-SFB-1073, project A05, and DFG-SPP-1840), the VolkswagenStiftung, and the European Union (EU) within the Horizon 2020 ERC-StG. "ULEED" (ID: 639119).

**References**

1. U. Bovensiepen, H. Petek, M. Wolf, *Dynamics at Solid State Surfaces and Interfaces* (Wiley-VCH Verlag GmbH & Co. KGaA, Weinheim, Germany, 2010), vol. 1.

2. R. J. D. Miller, *Annu. Rev. Phys. Chem.* **65**, 583–604 (2014).

3. A. H. Zewail, *Annu. Rev. Phys. Chem.* **57**, 65–103 (2006).

4. A. M. Lindenberg, *Science*. **308**, 392–395 (2005).

5. M. A. Huber, M. Plankl, M. Eisele, R. E. Marvel, F. Sandner, T. Korn, C. Schüller, R. F. Haglund, R. Huber, T. L. Cocker, *Nano Lett.* **16**, 1421–1427 (2016).

6. V. Kravtsov, R. Ulbricht, J. M. Atkin, M. B. Raschke, *Nat. Nanotechnol.* **11**, 1–7 (2016).

7. M. Wagner, Z. Fei, A. S. McLeod, A. S. Rodin, W. Bao, E. G. Iwinski, Z. Zhao, M. Goldflam, M. Liu, G. Dominguez, M. Thiemens, M. M. Fogler, A. H. Castro Neto, C. N. Lau, S. Amarie, F. Keilmann, D. N. Basov, *Nano Lett.* **14**, 894–900 (2014).

8. Y. Terada, S. Yoshida, O. Takeuchi, H. Shigekawa, *Nat. Photonics*. **4**, 869–874 (2010).

9. T. L. Cocker, V. Jelic, M. Gupta, S. J. Molesky, J. A. J. Burgess, G. D. L. Reyes, L. V. Titova, Y. Y. Tsui, M. R. Freeman, F. A. Hegmann, *Nat. Photonics*. **7**, 620–625 (2013).

10. T. L. Cocker, D. Peller, P. Yu, J. Repp, R. Huber, *Nature*. **539**, 263–267 (2016).

11. E. Quinonez, J. Handali, B. Barwick, *Rev. Sci. Instrum.* **84**, 103710 (2013).

12. M. Müller, A. Paarmann, R. Ernstorfer, *Nat. Commun.* **5**, 5292 (2014).

13. J. Vogelsang, J. Robin, B. J. Nagy, P. Dombi, D. Rosenkranz, M. Schiek, P. Groß, C. Lienau, *Nano Lett.* **15**, 4685–4691 (2015).

14. K. J. Gaffney, H. N. Chapman, *Science*. **316**, 1444–1448 (2007).

15. R. L. Sandberg, C. Song, P. W. Wachulak, D. A. Raymondson, A. Paul, B. Amirbekian, E. Lee, A. E. Sakdinawat, C. La-O-Vorakiat, M. C. Marconi, C. S. Menoni, M. M. Murnane, J. J. Rocca, H. C. Kapteyn, J. Miao, *Proc. Natl. Acad. Sci.* **105**, 24–27 (2008).

16. O. Kfir, S. Zayko, C. Nolte, M. Sivis, M. Möller, B. Hebler, S. S. P. K. Arekapudi, D. Steil, S. Schäfer, M. Albrecht, O. Cohen, S. Mathias, C. Ropers, *Sci. Adv.*, in press, doi:10.1126/sciadv.aao4641.

17. R. J. D. Miller, *Science*. **343**, 1108–1116 (2014).

18. A. H. Zewail, *Science*. **328**, 187–193 (2010).

19. P. Musumeci, J. T. Moody, C. M. Scoby, M. S. Gutierrez, H. A. Bender, N. S. Wilcox, *Rev. Sci. Instrum.* **81**, 13306 (2010).

20. S. Lahme, C. Kealhofer, F. Krausz, P. Baum, *Struct. Dyn.* **1**, 34303 (2014).

21. M. Gulde, S. Schweda, G. Storeck, M. Maiti, H. K. Yu, A. M. Wodtke, S. Schafer, C. Ropers, *Science*. **345**, 200–204 (2014).

22. S. P. Weathersby, G. Brown, M. Centurion, T. F. Chase, R. Coffee, J. Corbett, J. P. Eichner, J. C. Frisch, A. R. Fry, M. Gühr, N. Hartmann, C. Hast, R. Hettel, R. K. Jobe, E. N. Jongewaard, J. R. Lewandowski, R. K. Li, A. M. Lindenberg, I. Makasyuk, *et al.*, *Rev. Sci. Instrum.* **86**, 73702 (2015).

23. S. Manz, A. Casandruc, D. Zhang, Y. Zhong, R. A. Loch, A. Marx, T. Hasegawa, L. C. Liu, S. Bayesteh, H. Delsim-Hashemi, M. Hoffmann, M. Felber, M. Hachmann, F. Mayet, J. Hirscht, S. Keskin, M. Hada, S. W. Epp, K. Flöttmann, *et al.*, *Faraday Discuss.* **177**, 467–491 (2015).

24. W. Liang, S. Schäfer, A. H. Zewail, *Chem. Phys. Lett.* **542**, 1–7 (2012).

25. A. Hanisch-Blicharski, A. Janzen, B. Krenzer, S. Wall, F. Klasing, A. Kalus, T. Frigge, M. Kammler, M. Horn-von Hoegen, *Ultramicroscopy*. **127**, 2–8 (2013).

26. T. Frigge, B. Hafke, T. Witte, B. Krenzer, C. Streubühr, A. Samad Syed, V. Mikšić Trontl, I. Avigo, P. Zhou, M. Ligges, D. von der Linde, U. Bovensiepen, M. Horn-von Hoegen, S. Wippermann, A. Lücke, S. Sanna, U. Gerstmann, W. G. Schmidt, *Nature*. **544**, 207–211 (2017).

27. S. Vogelgesang, G. Storeck, J. G. Horstmann, T. Diekmann, M. Sivis, S. Schramm, K. Rossnagel, S. Schäfer, C. Ropers, *Nat. Phys.* **14**, 184–190 (2018).

28. H. Dömer, O. Bostanjoglo, *Rev. Sci. Instrum.* **74**, 4369 (2003).

29. J. S. Kim, T. LaGrange, B. W. Reed, M. L. Taheri, M. R. Armstrong, W. E. King, N. D. Browning, G. H. Campbell, *Science*. **321**, 1472–1475 (2008).

30. D.-S. Yang, O. F. Mohammed, A. H. Zewail, *Proc. Natl. Acad. Sci.* **107**, 14993–14998 (2010).

31. J. Sun, V. A. Melnikov, J. I. Khan, O. F. Mohammed, *J. Phys. Chem. Lett.* **6**, 3884–3890 (2015).

32. T. Van Oudheusden, E. F. De Jong, S. B. van der Geer, W. P. E. M. Op 'T Root, O. J. Luiten, B. J. Siwick, W. P. E. M. Op't Root, O. J. Luiten, B. J. Siwick, *J. Appl. Phys.* **102**, 93501 (2007).

33. G. Sciaini, R. J. D. Miller, *Reports Prog. Phys.* **74**, 96101 (2011).

34. R. P. Chatelain, V. R. Morrison, C. Godbout, B. J. Siwick, *Appl. Phys. Lett.* **101**, 81901 (2012).

35. C. Gerbig, A. Senftleben, S. Morgenstern, C. Sarpe, T. Baumert, *New J. Phys.* **17**, 43050 (2015).

36. L. Waldecker, R. Bertoni, R. Ernstorfer, *J. Appl. Phys.* **117**, 44903 (2015).

37. H. Daoud, K. Floettmann, R. J. D. Miller, *Struct. Dyn.* **4**, 0–9 (2017).

38. M. Reiser, *Theory and Design of Charged Particle Beams* (Wiley-VCH Verlag GmbH & Co. KGaA, Weinheim, Germany, 2008), *Wiley Series in Beam Physics and Accelerator Technology*.

39. D. H. Dowell, J. F. Schmerge, *Phys. Rev. Spec. Top. - Accel. Beams*. **12**, 74201 (2009).

40. M. Kuwahara, Y. Nambo, K. Aoki, K. Sameshima, X. Jin, T. Ujihara, H. Asano, K. Saitoh, Y. Takeda, N. Tanaka, *Appl. Phys. Lett.* **109**, 13108 (2016).

41. L. Kasmi, D. Kreier, M. Bradler, E. Riedle, P. Baum, *New J. Phys.* **17**, 33008 (2015).

42. M. Merano, S. Collin, P. Renucci, M. Gatri, S. Sonderegger, A. Crottini, J. D. Ganière, B. Deveaud, *Rev. Sci. Instrum.* **76**, 85108 (2005).

43. C. Ropers, D. R. Solli, C. P. Schulz, C. Lienau, T. Elsaesser, *Phys. Rev. Lett.* **98**, 43907 (2007).

44. B. Barwick, C. Corder, J. Strohaber, N. Chandler-Smith, C. Uiterwaal, H. Batelaan, *New J. Phys.* **9**, 142–142 (2007).

45. P. Hommelhoff, Y. Sortais, A. Aghajani-Talesh, M. A. Kasevich, *Phys. Rev. Lett.* **96**, 77401 (2006).

46. R. Bormann, M. Gulde, A. Weismann, S. V. Yalunin, C. Ropers, *Phys. Rev. Lett.* **105**, 147601 (2010).

47. G. Herink, D. R. Solli, M. Gulde, C. Ropers, *Nature*. **483**, 190–193 (2012).

48. D. Ehberger, J. Hammer, M. Eisele, M. Krüger, J. Noe, A. Högele, P. Hommelhoff, *Phys. Rev. Lett.* **114**, 227601 (2015).

49. A. Feist, N. Bach, N. Rubiano da Silva, T. Danz, M. Möller, K. E. Priebe, T. Domröse, J. G. Gatzmann, S. Rost, J. Schauss, S. Strauch, R. Bormann, M. Sivis, S. Schäfer, C. Ropers, *Ultramicroscopy*. **176**, 63–73 (2017).

50. F. Houdellier, G. M. Caruso, S. Weber, M. Kociak, A. Arbouet, *Ultramicroscopy*. **186**, 128–138 (2018).

51. A. Paarmann, M. Gulde, M. Müller, S. Schäfer, S. Schweda, M. Maiti, C. Xu, T. Hohage, F. Schenk, C. Ropers, R. Ernstorfer, *J. Appl. Phys.* **112**, 113109 (2012).

52. R. Bormann, S. Strauch, S. Schäfer, C. Ropers, *J. Appl. Phys.* **118**, 173105 (2015).

53. B. Schröder, M. Sivis, R. Bormann, S. Schäfer, C. Ropers, *Appl. Phys. Lett.* **107**, 231105 (2015).

54. M. Z. Hasan, C. L. Kane, *Rev. Mod. Phys.* **82**, 3045–3067 (2010).

55. A. Ichimiya, P. I. Cohen, *Reflection High-Energy Electron Diffraction* (Cambridge University Press, Cambridge, 2004), vol. 6 of *Springer Series in Surface Sciences*.

56. M. A. Van Hove, W. H. Weinberg, C.-M. Chan, *Low-Energy Electron Diffraction* (Springer Berlin Heidelberg, Berlin, Heidelberg, 1986), vol. 6 of *Springer Series in Surface Sciences*.

57. G. Storeck, S. Vogelgesang, M. Sivis, S. Schäfer, C. Ropers, *Struct. Dyn.* **4**, 44024 (2017).

58. A. Spijkerman, J. L. de Boer, A. Meetsma, G. A. Wiegers, S. van Smaalen, *Phys. Rev. B*. **56**, 13757–13767 (1997).

59. K. Rossnagel, *J. Phys. Condens. Matter.* **23**, 213001 (2011).

60. L. Stojchevska, I. Vaskivskyi, T. Mertelj, P. Kusar, D. Svetin, S. Brazovskii, D. Mihailovic, *Science*. **344**, 177–180 (2014).

61. L. Perfetti, P. A. Loukakos, M. Lisowski, U. Bovensiepen, H. Berger, S. Biermann, P. S. Cornaglia, A. Georges, M. Wolf, *Phys. Rev. Lett.* **97**, 67402 (2006).

62. S. Hellmann, C. Sohrt, M. Beye, T. Rohwer, F. Sorgenfrei, M. Marczynski-Bühlow, M. Kalläne, H. Redlin, F. Hennies, M. Bauer, A. Föhlisch, L. Kipp, W. Wurth, K. Rossnagel, *New J. Phys.* **14**, 13062 (2012).

63. J. C. Petersen, S. Kaiser, N. Dean, A. Simoncig, H. Y. Liu, A. L. Cavalieri, C. Cacho, I. C. E. Turcu, E. Springate, F. Frassetto, L. Poletto, S. S. Dhesi, H. Berger, A. Cavalleri, *Phys. Rev. Lett.* **107**, 177402 (2011).

64. M. Ligges, I. Avigo, D. Golež, H. Strand, L. Stojchevska, M. Kalläne, P. Zhou, K. Rossnagel, M. Eckstein, P. Werner, U. Bovensiepen, (2017) (available at http://arxiv.org/abs/1702.05300).

65. M. Eichberger, H. Schäfer, M. Krumova, M. Beyer, J. Demsar, H. Berger, G. Moriena, G. Sciaini, R. J. D. Miller, *Nature*. **468**, 799–802 (2010).

66. K. Haupt, M. Eichberger, N. Erasmus, A. Rohwer, J. Demsar, K. Rossnagel, H. Schwoerer, *Phys. Rev. Lett.* **116**, 16402 (2016).

67. C. Laulhé, T. Huber, G. Lantz, A. Ferrer, S. O. Mariager, S. Grübel, J. Rittmann, J. A. Johnson, V. Esposito, A. Lübcke, L. Huber, M. Kubli, M. Savoini, V. L. R. Jacques, L. Cario, B. Corraze, E. Janod, G. Ingold, P. Beaud, *et al.*, *Phys. Rev. Lett.* **118**, 247401 (2017).

68. W. L. McMillan, *Phys. Rev. B*. **12**, 1187–1196 (1975).

69. L. Wimmer, G. Herink, D. R. Solli, S. V Yalunin, K. E. Echternkamp, C. Ropers, *Nat. Phys.* **10**, 432–436 (2014).

70. C. Kealhofer, W. Schneider, D. Ehberger, A. Ryabov, F. Krausz, P. Baum, *Science*. **352**, 429–433 (2016).

71. D. J. Flannigan, A. H. Zewail, *Acc. Chem. Res.* **45**, 1828–1839 (2012).

72. L. Piazza, D. J. Masiel, T. LaGrange, B. W. Reed, B. Barwick, F. Carbone, *Chem. Phys.* **423**, 79–84 (2013).

73. A. Feist, K. E. Echternkamp, J. Schauss, S. V. Yalunin, S. Schäfer, C. Ropers, *Nature*. **521**, 200–203 (2015).

74. E. Kieft, K. B. Schliep, P. K. Suri, D. J. Flannigan, *Struct. Dyn.* **2**, 51101 (2015).

75. G. Cao, S. Sun, Z. Li, H. Tian, H. Yang, J. Li, *Sci. Rep.* **5**, 8404 (2015).

76. K. Bücker, M. Picher, O. Crégut, T. LaGrange, B. W. Reed, S. T. Park, D. J. Masiel, F. Banhart, *Ultramicroscopy*. **171**, 8–18 (2016).

77. Y. M. Lee, Y. J. Kim, Y.-J. Kim, O.-H. Kwon, *Struct. Dyn.* **4**, 44023 (2017).

78. S. Ji, L. Piazza, G. Cao, S. T. Park, B. W. Reed, D. J. Masiel, J. Weissenrieder, *Struct. Dyn.* **4**, 54303 (2017).

79. W. Verhoeven, J. F. M. van Rens, E. R. Kieft, P. H. A. Mutsaers, O. J. Luiten, *Ultramicroscopy*. **188**, 85–89 (2018).

80. A. Tonomura, *Electron Holography* (Springer Berlin Heidelberg, Berlin, Heidelberg, ed. 2, 1999), vol. 70 of *Springer Series in Optical Sciences*.

81. P. A. Midgley, R. E. Dunin-Borkowski, *Nat. Mater.* **8**, 271–280 (2009).

82. J. Verbeeck, H. Tian, P. Schattschneider, *Nature*. **467**, 301–304 (2010).

83. J. M. Zuo, J. C. H. Spence, *Advanced Transmission Electron Microscopy* (Springer New York, New York, NY, 2017; http://link.springer.com/10.1007/978-1-4939-6607-3).

84. B. Cook, M. Bronsgeest, K. Hagen, P. Kruit, *Ultramicroscopy*. **109**, 403–412 (2009).

85. D. K. Armani, T. J. Kippenberg, S. M. Spillane, K. J. Vahala, *Nature*. **421**, 925–928 (2003).

86. M. Eichenfield, J. Chan, R. M. Camacho, K. J. Vahala, O. Painter, *Nature*. **462**, 78–82 (2009).

87. M. Aspelmeyer, T. J. Kippenberg, F. Marquardt, *Rev. Mod. Phys.* **86**, 1391–1452 (2014).

88. M. Maldovan, *Nature*. **503**, 209–217 (2013).

89. S. Volz, J. Ordonez-Miranda, A. Shchepetov, M. Prunnila, J. Ahopelto, T. Pezeril, G. Vaudel, V. Gusev, P. Ruello, E. M. Weig, M. Schubert, M. Hettich, M. Grossman, T. Dekorsy, F. Alzina, B. Graczykowski, E. Chavez-Angel, J. Sebastian Reparaz, M. R. Wagner, *et al.*, *Eur. Phys. J. B*. **89**, 15 (2016).

90. B. Barwick, H. S. Park, O. Kwon, J. S. Baskin, A. H. Zewail, *Science*. **322**, 1227–1231 (2008).

91. A. Yurtsever, A. H. Zewail, *Proc. Natl. Acad. Sci.* **108**, 3152–3156 (2011).

92. D. R. Cremons, D. A. Plemmons, D. J. Flannigan, *Nat. Commun.* **7**, 11230 (2016).

93. A. Feist, N. Rubiano da Silva, W. Liang, C. Ropers, S. Schäfer, *Struct. Dyn.* **5**, 14302 (2018).

94. K. E. Echternkamp, A. Feist, S. Schäfer, C. Ropers, *Nat. Phys.* **12**, 1000–1004 (2016).

95. K. E. Priebe, C. Rathje, S. V. Yalunin, T. Hohage, A. Feist, S. Schäfer, C. Ropers, *Nat. Photonics*. **11**, 793–797 (2017).

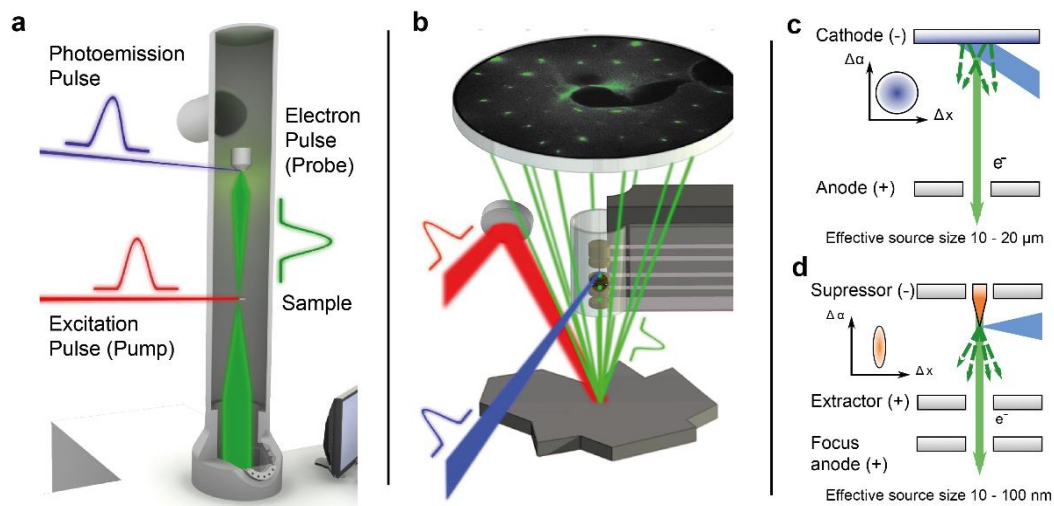

**Figure 1.** Ultrafast electron microscopy and diffraction using laser-triggered field emitters. Optically induced dynamics are probed (a) in transmission by UTEM (electron kinetic energy: 120-200 keV) and (b) at surfaces by ULEED (electron kinetic energy: 10-500 V), employing low-emittance ultrashort electron pulses. (c,d) Conceptual design of ultrafast photoemission electron sources. In a planar photocathode geometry (c), electrons are emitted from micrometer-sized areas (governed by typical laser focal spot diameters of 10-20 µm), while the laser-triggered field emitter (d) localizes electron emission to the nanoscale apex of a sharp metal tip (typical diameter of 10-100 nm), facilitating high-coherence electron pulses. Insets: Illustration of occupied transverse phase space for the respective emitter geometry. For details of panel (a,c) and (b) see Ref. *49* and Ref. *57*, respectively.

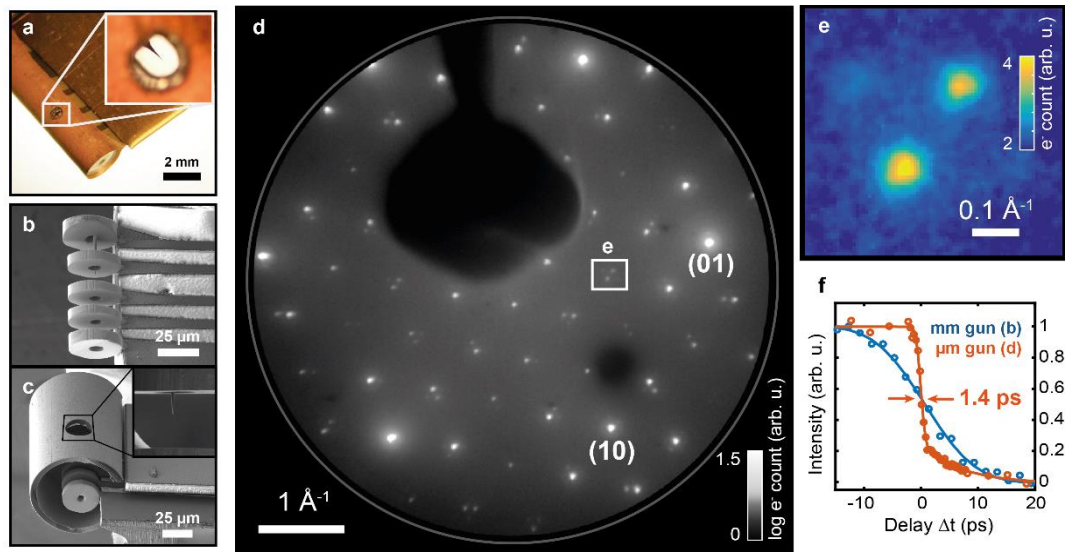

**Figure 2.** Laser-driven electron guns for ULEED. (a) Millimeter-scale electrostatic lens assembly with a nanotip field emitter (optical microscope image). (b) Micrometer-scale electron gun, featuring Schottky-type extractor-suppressor unit and collimation einzel lens (scanning electron micrograph). (c) Electron gun covered by ground shield with hole for laser excitation. Inset: Nanotip photocathode. (d) ULEED diffraction pattern from a 1T-TaS2 surface recorded with photoelectrons (room temperature, 100 eV energy, millimeter-scale gun), showing lattice Bragg peaks (some labeled) and satellite peaks from a periodic lattice distortion. (e) Zoom-in of region denoted in (d), resolving adjacent higher-order satellite peaks. (f) Temporal resolution of electron guns measured by laser-induced change in diffraction intensity (cf. Fig. 3). (a,d-f): Reproduced with permission from Ref. *27*. DOI: 10.1038/nphys4309. © 2017 Macmillan Publishers Limited, part of Springer Nature. All rights reserved. (b,c) For details see Ref. *57*.

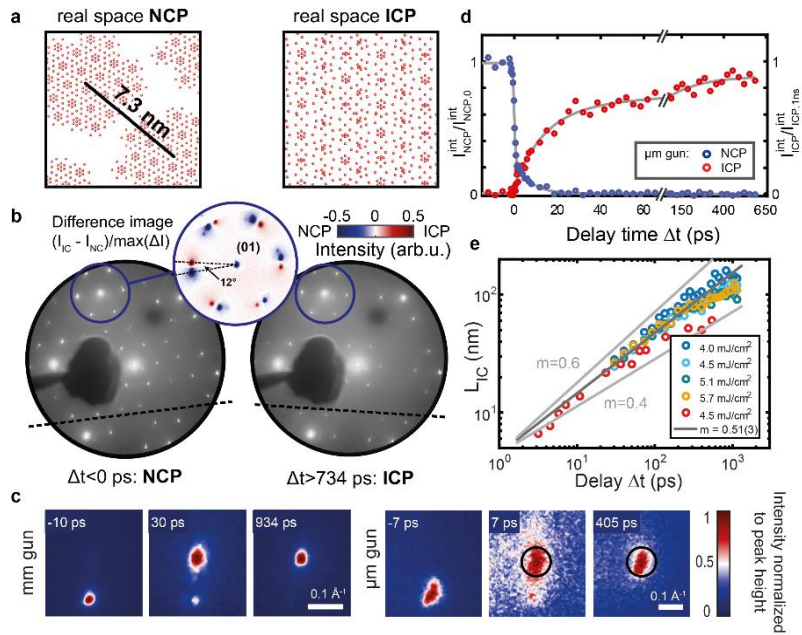

**Figure 3.** Structural phase transition and phase-ordering kinetics probed by ULEED. (a) Sketches of the periodic lattice distortion in 1T-TaS$_2$ corresponding to the charge-density wave in the domain-like "nearly commensurate" (NCP, left) and incommensurate (ICP, right) phase. White areas in the NCP denote discommensurations with weak lattice distortion (cf. Ref. *58*). Distortions of the underlying hexagonal lattice strongly exaggerated. (b) Diffraction images before (left) and after (right) pump pulse; zoom-in of difference image, indicating the appearance of rotated ICP spots. (c) Delay-dependent maps of individual diffraction spots, showing NCP-to-ICP switching and the time-dependent ICP spot narrowing. (d) Delay-dependent suppression of NCP and appearance of ICP diffraction intensity. (e) Universal scaling of the phase-ordering kinetics in the ICP for different excitation fluences, evidenced by the power-law growth of the spatial correlation length (red: µm-sized gun; yellow to blue: mm-sized gun). Adapted with permission from Ref. *27*. DOI: 10.1038/nphys4309. © 2017 Macmillan Publishers Limited, part of Springer Nature. All rights reserved.

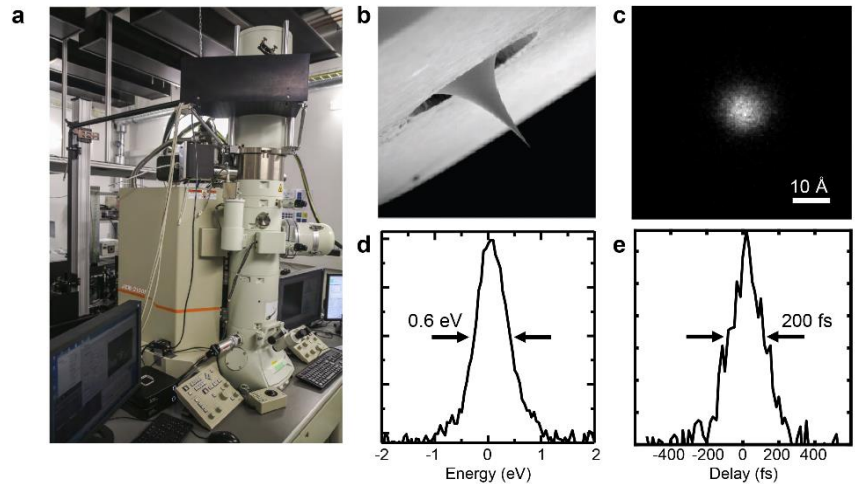

**Figure 4.** Experimental setup and selected electron pulse properties of the Göttingen UTEM instrument. (a) A laser-driven Schottky field emission electron gun is integrated with a custom modified JEOL JEM-2100F. (b) Nanoscale photocathodes facilitate the generation of ultrafast electron pulses with (c-e) down to 0.89-nm focal spot size, 0.6-eV energy width and 200-fs pulse duration. For details see Ref. *49*.

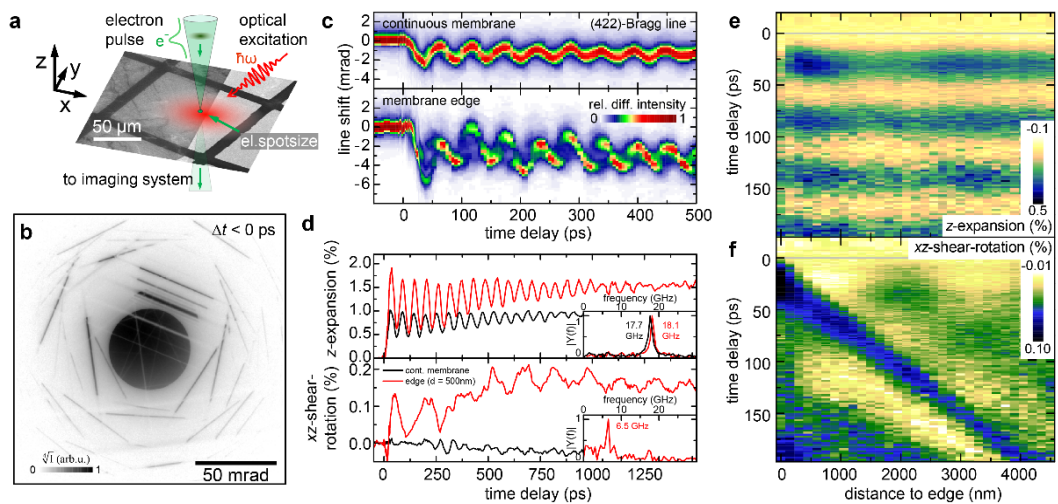

**Figure 5.** Ultrafast nanoscale diffractive probing of strain dynamics. (a) A laser-excited single graphite thin film is probed close to its edge with a sharply focused electron beam (28-nm focal spot size). (b) Ultrafast convergent beam electron diffraction (U-CBED) pattern. (c) Time-dependent profiles of the (422)-Bragg line recorded in a continuous part of the membrane and close to its edge (500-nm distance). (d) Quantitative reconstruction of local deformation gradient tensor components, disentangling the involved mechanical deformation modes. (e) Distance- and time delay dependent strain dynamics reveal the complex spatio-temporal strain dynamics. For details see Ref. *93*.

17